\documentclass[twocolumn]{aastex631}

\shorttitle{Secondary electron spectrum in CR discharge}
\shortauthors{Yutaka Ohira}
\usepackage{natbib}
\begin{document}

\title{Evolution of secondary electron spectrum during cosmic-ray discharge in the universe}
\author[0000-0002-2387-0151]{Yutaka Ohira}
\affiliation{Department of Earth and Planetary Science, The University of Tokyo, \\
7-3-1 Hongo, Bunkyo-ku, Tokyo 113-0033, Japan}
\email{y.ohira@eps.s.u-tokyo.ac.jp}

\begin{abstract}
We recently found that streaming cosmic rays (CRs) induce a resistive electric field that can accelerate secondary electrons produced by CR ionization. 
In this work, we study the evolution of the energy spectrum of secondary electrons by numerically solving the one-dimensional Boltzmann equation and Ohm's law. 
We show that the accelerated secondary electrons further ionize a gas, that is, the electron avalanche occurs, resulting in increased ionization and excitation of the gas. 
Although the resistive electric field becomes weaker than one before the CR discharge, the weak resistive electric field weakly accelerates the secondary electrons. 
The quasi-steady state is almost independent of the initial resistive electric field, but depends on the electron fraction in the gas. 
The resistive electric field in the quasi-steady state is larger for the higher electron fraction, 
which makes the number of secondary electrons that can ionize the gas larger, resulting in a higher ionization rate. 
The CR discharge could explain the high ionization rate that are observed in some molecular clouds. 
\end{abstract}

\section{Introduction}
\label{sec1}
In the present universe, there are nonthermal particles with energies much higher than the energy scale of the system, which are called cosmic rays (CRs). 
CRs can easily ionize neutral gases in the universe \citep{hayakawa61,spitzer68}. 
The ionization rate, $\zeta=n_{\rm H}^{-1}dn_{\rm H}/dt$, was originally estimated to be about $10^{-17} {\rm s}^{-1}$ based on the CR flux observed at Earth. 
However, observations of emission and absorption lines from ionized molecules sometimes suggest ionization rates one or two orders of magnitude higher than the expectation  \citep{caselli98, indriolo12,indriolo15}. 
A part of those high ionization rates can be explained by inhomogeneity of low energy Galactic CRs that mainly contribute the ionization rate \citep{phan23}. 
To explain high ionization rates in high density regions, some particle accelerations in the dense molecular clouds were proposed \citep{padovani16,yang20,gaches21}. 
Comprehensive reviews of the CR ionization in molecular clouds are presented by \citet{dalgarno06,padovani20,gabici22}. 

In the standard picture of CR ionization, secondary electrons generated by CR ionization, whose energy is typically about $10 ~{\rm eV}$, lose their energy through further ionization, excitation, and Coulomb scattering. 
On average, one secondary electron ionizes about one hydrogen molecule during the energy loss process \citep{ivlev21}.
We recently proposed that the secondary electrons can be accelerated in self-discharge by streaming CRs, so that ionization by the secondary electrons is enhanced \citep{ohira22}. 
The streaming CR protons generate the CR electric current, so that the return current of background electrons is induced to compensate the CR current, that is, the background electrons have a drift velocity with respect to the background protons and hydrogen atoms. 
Then, the resistive electric field is induced because there is a finite resistivity between the background electrons and protons (or hydrogen atoms). 
As a result, secondary electrons can be accelerated by the resistive electric field as long as the CR current is sufficiently large. 
The resistive electric field induced by CRs was first discussed in \citet{malkov16} to explain the CR positron anomaly. 
Moreover, the authors mentioned the discharge by the resistive electric field in \citet{malkov16}.  
In our previous work \citep{ohira22}, we derive the discharge condition by assuming that all secondary electrons have the same energy of $10~{\rm eV}$. 
In addition we solve the time evolution of the resistive electric field during the discharge by using a fluid approximation for the secondary electrons and neglecting the friction force acting on the secondary electron fluid. 

In this work, to understand the discharge process by the streaming CR more deeply, we investigate the spatial evolution of the resistive electric field and the energy spectrum of secondary electrons by solving the Boltzmann equation of the secondary electrons. 
We first estimate the streaming CR flux around a SNR in Section \ref{sec2}. 
Then, we give the equation system that we solve in this work in Section \ref{sec3}. 
In section \ref{sec4} , we show the results of numerical calculations. 
Finally, we discuss the prospects for future work in Section  \ref{sec5} and summarize this work in Section \ref{sec6}.

\section{CR flux around a SNR}
\label{sec2} 
In this section, we estimate the streaming CR flux around a SNR to estimate the resistive electric field induced by the streaming CRs. 
Since there are turbulent magnetic fields in the current galaxy, the momentum distribution of Galactic CRs is almost isotropic unless in the vicinity of a CR source. 
In order to explain the observed Galactic CR data at the GeV scale, the diffusive flux of CR protons escaping from 
a SNR has to be about 100 times larger than that of electrons. 
We therefore consider only protons as CRs in this work although CR electrons may contribute to the CR flux when we consider lower energy CRs \citep{silsbee20}. 
In the standard model of Galactic CRs, one SNR produces on average $10^{50}~{\rm erg}$ in CR protons, that is, the total number of CRs with energy of $1~{\rm GeV}$ is $N_{\rm CR}\approx 6\times 10^{52}$. 
Since CRs diffuse to surroundings after escaping from the SNR, the escaping CRs have the diffusive flux which corresponds to the streaming CR flux. 
The diffusion length is given by
\begin{eqnarray}
R_{\rm diff}&=& \sqrt{4\pi D t} \nonumber \\
&=&20~{\rm pc} \left(\frac{D}{10^{28}~{\rm cm^2~s^{-1}}}\right)^{\frac{1}{2}}\left(\frac{t}{1~{\rm kyr}}\right)^{\frac{1}{2}}~~,
\label{eq:diff}
\end{eqnarray}
where $D$ and $t$ are the diffusion coefficient of CRs and elapsed time since the CRs left the SNR. 
The mean CR density within the diffusion length is $n_{\rm CR}\sim N_{\rm CR}/R_{\rm diff}^3$ for the isotropic diffusion. 
Then, the diffusive flux density at the point of the diffusion length away from the SNR is roughly estimated by
\begin{eqnarray}
F_{\rm CR, iso} &\sim& D\frac{n_{\rm CR}}{R_{\rm diff}} \nonumber \\
                 &=&44~{\rm cm^2~s^{-1}} \left(\frac{D}{10^{28}~{\rm cm^2~s^{-1}}}\right)^{-1}\left(\frac{t}{1~{\rm kyr}}\right)^{-2}~~, 
\label{eq:fcr1}
\end{eqnarray}
Gamma-ray observations and theoretical studies suggest that the diffusion coefficient around SNRs are of the order of $10^{-26}~{\rm cm^2~s^{-1}}$, which is about 100 times smaller than the value expected from studies for CR propagation in our galaxy \citep{torres08,fujita09,fujita10}. 
Therefore, the diffusive flux in the diffusion length is expected to be 100 times larger than the above estimate, but the diffusion length is 10 times smaller than the estimate in Equation (\ref{eq:diff}). 
In such cases, CRs cannot propagate far from the SNR and the CR distribution should be estimated in a more accurate way \citep{ohira11}.

If CRs preferentially diffuse along the large scale magnetic field line \citep{malkov13}, the mean CR density is $n_{\rm CR}\sim N_{\rm CR}/R_{\rm diff}R_{\rm SNR}^2$, where all CRs are assumed to be produced mostly when the SNR radius is $R_{\rm SNR}$. Then, the diffusive flux along the magnetic field line is estimated by 
\begin{eqnarray}
F_{\rm CR, 1D} &\sim& D\frac{n_{\rm CR}}{R_{\rm diff}} \nonumber \\
                 &=&184~{\rm cm^2~s^{-1}} \left(\frac{R_{\rm SNR}}{10~{\rm pc}}\right)^{-2}\left(\frac{t}{1~{\rm kyr}}\right)^{-1}~~.  
\label{eq:fcr2}
\end{eqnarray}
%
In this case, the diffusive flux does not depend on the diffusion coefficient. 

The resistive electric field is induced by the CR flux and 
the condition for secondary electrons with $10~{\rm eV}$ to be accelerated by the resistive electric field is given in \citet{ohira22}: 
\begin{equation}
F_{\rm CR} > 32~{\rm cm^{-2}~ s^{-1}} \left(\frac{T_{\rm e}}{10~{\rm K}}\right)^{\frac{3}{2}} \left(\frac{n_{\rm H}}{100~{\rm cm}^{-3}}\right) 
\label{eq:dc}
\end{equation}
for $n_{\rm e}/n_{\rm H} \le 8.3\times 10^{-4}$, where $T_{\rm e}, n_{\rm e}$ and $n_{\rm H}$ are the electron temperature, number densities of free electrons and hydrogen atoms, respectively. 
It should be noted that secondary electrons with higher energies than $10~{\rm eV}$ can be produced by CR ionization. 
These higher energy secondary electrons can be accelerated with a CR flux lower than the above condition. 
For this reason, the above discharge condition should only be used as a reference.  
As estimated in the above, the CR flux density around a SNR is about $10-10^3~{\rm cm^2~s^{-1}}$. 
Therefore, the CR discharge is expected to occur in a cold HI or molecular cloud around a SNR. 

For the isotropic and one-dimensional diffusions, the mean streaming velocity of CRs around the SNR is estimated to be 
\begin{eqnarray}
V_{\rm CR} &\sim&\frac{F_{\rm CR}}{n_{\rm CR}} =\frac{D}{R_{\rm diff}}\nonumber \\
&=&1.6 \times 10^8~{\rm cm~ s^{-1}}  \left(\frac{D}{10^{28}~{\rm cm^2~s^{-1}}}\right)^{\frac{1}{2}}\left(\frac{t}{1~{\rm kyr}}\right)^{-\frac{1}{2}}
\label{eq:vcr}
\end{eqnarray}
%

\section{Boltzmann equation and Ohm's law}
\label{sec3} 

Secondary electrons are accelerated by the resistive electric field as long as the CR flux is sufficiently large. 
The accelerated secondary electrons generate additional secondary electrons. 
Then, the current of secondary electrons increases, but the current of thermal electrons decreases to satisfy the current neutrality condition. 
The resistivity of secondary electron current is smaller than that of thermal electrons because the secondary electrons have large energy compared with thermal electrons. 
As a result, the strength of the resistive electric field decreases as the discharge goes on.

We analytically solved the time evolution of the CR discharge by using many assumptions and approximations in our previous work \citep{ohira22}. 
In particular, we did not consider ionization by secondary electrons, that is, the electron avalanche process was not included. 
In this work, we numerically solve the steady-state one-dimensional spatial evolution of energy spectrum of secondary electrons in the direction of the CR streaming, where ionization by primary CRs and secondary electrons is considered. 
The velocity space is also restricted to the one dimension in the direction of CR streaming and the nonrelativistic limit, $0< v_{\rm x}\ll c$. 
The kinetic energy, $\varepsilon=m_{\rm e} v_{\rm x}^2/2$, is used instead of the velocity in this work.

The Boltzmann equation for secondary electrons in the steady state is given by
\begin{equation}
v_{\rm x}\frac{\partial f}{\partial x} +\frac{\partial }{\partial \varepsilon} \left \{ \left( \frac{d\varepsilon}{dt} \right) f\right \} = Q_{\rm i,CR}+Q_{\rm i,2nd}+Q_{\rm ex,2nd},
\label{eq:boltzmann}
\end{equation}
where $f(x,\varepsilon)$ is the phase space density in the one dimensional system, 
and $d{\varepsilon}/dt$ describes the energy gain by the resistive electric field and energy loss due to the Coulomb interaction. 
Instead of solving the above equation for $f$, we define $h \equiv v_{\rm x} f$ and solve the following equation for $h$:
\begin{equation}
\frac{\partial h}{\partial x} + \frac{\partial }{\partial \varepsilon} \left \{ \left( \frac{d\varepsilon}{dx} \right) h\right \} = Q_{\rm i,CR}+Q_{\rm i,2nd}+Q_{\rm ex,2nd}.
\label{eq:boltzmann2}
\end{equation}
The first and second terms on the left hand side describe advection in the one dimensional spatial space and energy space, respectively. 
$d\varepsilon/dx$ is given by 
\begin{equation}
\frac{d\varepsilon}{dx}=-eE_{\rm x}-\left(\frac{d\varepsilon}{dx}\right)_{\rm C}
\end{equation}
The first and second terms describe acceleration by the resistive electric field and the Coulomb loss, respectively. 
In this work, we use the continuous slowing-down approximation for the Coulomb loss, 
but exactly solve energy loss processes for ionization and excitation as discrete processes, which are described by the second and third terms on the right hand side of Equation~(\ref{eq:boltzmann2}). 
The energy loss rate due to the Coulomb loss, $(d\varepsilon/dx)_{\rm C}$, is given by \citet{swartz71}: 
\begin{equation}
\left(\frac{d\varepsilon}{dx}\right)_{\rm C} = -3.37\times 10^{-12}~{\rm eV~cm^{-1}} \left( \frac{n_{\rm e}}{1~{\rm cm}^{-3}} \right) \left(\frac{\varepsilon}{1~{\rm eV}}\right)^{-0.94}  . 
\end{equation}
The resistive electric field is given by Ohm's law:
\begin{equation}
E_{\rm x}=-\frac{ m_{\rm e}V_{\rm e}}{et_{\rm C}}~~,
\end{equation}
where $m_{\rm e}, V_{\rm e}$, and $e$, are the electron mass, mean background electron velocity with respect to background protons, and elementary charge, respectively. $t_{\rm C}$ is timescale for the Coulomb scattering between electrons and protons and given by 
\begin{equation}
t_{\rm C}=1.9\times10^{-1}~{\rm sec} \left(\frac{n_{\rm e}}{1~{\rm cm}^{-3}}\right)^{-1} \left(\frac{T_{\rm e}}{10~{\rm K}}\right)^{\frac{3}{2}}.
\label{eq:tc}
\end{equation}
From the current neutrality condition ($en_{\rm e}V_{\rm e} = J_{\rm CR} + J_{\rm 2nd}$), the resistive electric field is represented by the current densities of CRs and secondary electrons, $J_{\rm CR}=eF_{\rm CR}$ and $J_{\rm 2nd}$: 
\begin{equation}
E_{\rm x} = -\frac{m_{\rm e}}{e^2n_{\rm e} t_{\rm C}} \left( J_{\rm CR} + J_{\rm 2nd} \right), 
\end{equation}
where the current of the secondary electrons is calculated by 
\begin{equation}
J_{\rm 2nd} = -e \int h(x,\varepsilon) d\varepsilon. 
\end{equation}

The three terms on the right hand side of Equation~(\ref{eq:boltzmann2}) ($Q_{\rm i,CR}$, $Q_{\rm i,2nd}$, and  $Q_{\rm ex,2nd}$) are given by
\begin{eqnarray}
Q_{\rm i,CR}  &=& n_{\rm H}\int \frac{d\sigma_{\rm i,p}}{d\varepsilon}(\varepsilon,\varepsilon_0) vf_{\rm CR}(\varepsilon_0)d\varepsilon_0 ,\\
Q_{\rm i,2nd} &=& n_{\rm H} \left \{ \int \frac{d\sigma_{\rm i,e}}{d\varepsilon}(\varepsilon,\varepsilon_0) h(\varepsilon_0)d\varepsilon_0  -\sigma_{\rm i,e}(\varepsilon)h(\varepsilon) \right\} ,\\
Q_{\rm ex,2nd}  &=& n_{\rm H} \left\{ \sum_{k}  \sigma_{{\rm ex},k}(\varepsilon+\Delta_k)h(\varepsilon+\Delta_k)- \sigma_{{\rm ex},k}(\varepsilon)h(\varepsilon) \right \}, \nonumber \\
\end{eqnarray}
where $v$ and $f_{\rm CR}$ are the velocity corresponding to the kinetic energy of $\varepsilon_0$ and energy spectrum of CRs. 
$d\sigma_{\rm i,p}/d\varepsilon$ and $d\sigma_{\rm i,e}/d\varepsilon$ are  the differential cross sections for ionization by protons and electrons, respectively. 
$\sigma_{\rm i,e}$ and $\sigma_{{\rm ex},k}$ are the total cross sections for ionization and excitation by electrons, respectively. 
The subscript $k$ represents some excitation processes for a hydrogen atom or molecule, and $\Delta_k$ is the corresponding excitation energy. 
In this work, we consider thirteen excitation processes for a hydrogen molecule: vibrational transition ($v = 0 \rightarrow 1$), electronic excitation to singlet states ($B^1\Sigma^{+}_u, B'^1\Sigma^{+}_u, B''^1\Sigma^{+}_u, C^1\Pi_u , D^1\Pi_u , D'^1\Pi_u , EF^1\Sigma^{+}_{g}$), electronic excitation to triplet states ($a^3\Sigma^{+}_g , b^3\Sigma^{+}_u, c^3\Pi_u,$  $e^3\Sigma^{+}_u, d^3\Pi_u$). 
$Q_{\rm i,CR}$ describes injection of secondary electrons by primary CR ionization. 
The first term of $Q_{\rm i,2nd}$ describes injection of two types of secondary electrons with $\varepsilon$ by secondary electron impact ionization, new secondary electrons and secondary electrons that have lost their energy due to the ionization. 
The second term of $Q_{\rm i,2nd}$ describes loss of secondary electrons with $\varepsilon$ by secondary electron impact ionization. 
Similarly to $Q_{\rm i,2nd}$, the first and second terms of $Q_{\rm ex,2nd}$ describe injection and loss of secondary electrons with $\varepsilon$ by excitation, but new secondary electrons are not generated in the excitation process.

In this work, we do not solve the spatial evolution of primary CR spectrum and physical state of the background gas, but assume spatially uniform distributions, 
that is, we simply assume a constant CR current and a uniform background gas. 
This approximation for primary CRs is valid as long as the energy loss of primary CRs is negligible and the dynamical timescale of the CR source is longer than the timescale that we consider here. 
For simplicity, we use the mono-energetic spectrum of primary CRs, $f_{\rm CR}=n_{\rm CR}\delta(\varepsilon-m_{\rm p}c^2)$ in this work, where $n_{\rm CR}$ is the number density of primary CRs. 
Then, $Q_{\rm i,CR}$ is represented by 
\begin{equation}
Q_{\rm i,CR} (\varepsilon) = n_{\rm H}n_{\rm CR} \left\{ v\frac{d\sigma_{\rm i,p}}{d\varepsilon}(\varepsilon,\varepsilon_0)  \right\}_{\varepsilon_0=m_{\rm p}c^2}.
\label{eq:qcr}
\end{equation}
The resistivity heats gas \citep{miniati11,yokoyama23} and the back reaction of CR discharge on the background gas would change the temperature and ionization fraction. 
Since these are an interesting point to study, we should address this point in future.

We set the left boundary at $x=0$ and solve the spatial development of secondary electron spectrum in $x>0$. 
It should be noted that the directions of CR propagation and development of discharge are assumed to be the positive x direction in this work. 
The region in $x<0$ is assumed to be warm or hot medium, in which the resistivity and the resistive electric field are negligible. 
On the other hand, the region in $x\geq0$ is assumed to be a cold medium (molecular cloud). 
Then, discharge processes start to happen at $x=0$ and a quasi-steady state is expected to be realized at a sufficient distance from $x=0$.

Parameters to characterize the cold region are the temperature, $T$, hydrogen density, $n_{\rm H}$, and electron density, $n_{\rm e}$. 
Parameters to characterize primary CRs are the CR density, $n_{\rm CR}$ and mean drift velocity of CRs, $V_{\rm CR}$.
Hence, there are five parameters in this equation system. 
However, key parameters can be reduced to three by an appropriate normalization of this equation system. 
In the cold region, there are two characteristic length scales, the ionization length, 
\begin{equation}
l_{\rm ion}=\frac{1}{4\pi a_0^2 n_{\rm H}}~, 
\label{eq:lion}
\end{equation}
and the acceleration length, 
\begin{equation}
l_{\rm acc}=\frac{I_{\rm ion}}{eE_{\rm x,0}}~, 
\label{eq:lacc}
\end{equation}
where $a_0=5.29\times10^{-9}~{\rm cm}, I_{\rm ion}$ and $E_{\rm x,0}=m_{\rm e}n_{\rm CR}V_{\rm CR}/en_{\rm e}t_{\rm C}$ are the Bohr radius, first ionization potential of hydrogen atom or molecule, and resistive electric field before the CR discharge occurs ($E_{\rm x}$ at $x=0$). 
Secondary electrons with energy about $I_{\rm ion}$ ionize on average one hydrogen atom or molecule and lose their energy by $I_{\rm ion}$ in the ionization length, $l_{\rm ion}$. 
On the other hand, secondary electrons gain the energy of $I_{\rm ion}$ by the initial resistive electric field in the acceleration length, $l_{\rm acc}$. 
Hence, for $l_{\rm ion} /l_{\rm acc} > 1$, the secondary electrons can be accelerated by the resistive electric field before they lose their energy though ionization. 
The condition,  $l_{\rm ion} /l_{\rm acc} >1$, corresponds to the discharge condition given in Equation~(\ref{eq:dc}), but the two conditions do not exactly the same because the cross section of electron impact ionization is slightly different from $4\pi a_0^2$. 
The precise discharge condition, which corresponds to Equation~(\ref{eq:dc}), is given by $l_{\rm ion} /l_{\rm acc} > 0.167$.
Using Equations (\ref{eq:tc}), (\ref{eq:lion}) and (\ref{eq:lacc}), $l_{\rm ion} /l_{\rm acc}$ can be related to the cosmic-ray flux density by the following equation, 
\begin{equation}
\frac{l_{\rm ion}}{l_{\rm acc}} = 0.55\left(\frac{F_{\rm CR}}{10^2~{\rm cm^{-2}~ s^{-1}}} \right) \left(\frac{T_{\rm e}}{10~{\rm K}}\right)^{-\frac{3}{2}} \left(\frac{n_{\rm H}}{100~{\rm cm}^{-3}}\right)^{-1} 
\end{equation}
As estimated in Section~\ref{sec2}, the discharge condition can be satisfied in a cold atomic or molecular cloud around a SNR. If the gas temperature is $100~{\rm K}$, a large CR flux on the order of $10^3~{\rm cm^{-2}~ s^{-1}}$ is required to satisfy the discharge condition. 

By introducing new dimensionless variables, $\tilde{x}=x/l_{\rm ion}, \tilde{\varepsilon}=\varepsilon/I_{\rm ion}, \tilde{h}=hI_{\rm ion}/n_{\rm CR}c$, and $\tilde{E}_{\rm x}=E_{\rm x}/E_{\rm x,0}$, Equations~(\ref{eq:boltzmann2}) - (\ref{eq:qcr}) can be rewritten as follows:
\begin{eqnarray}
\frac{\partial \tilde{h}}{\partial \tilde{x}} &+&\frac{\partial }{\partial \tilde{\varepsilon}} \left \{ \left(-\frac{l_{\rm ion}}{l_{\rm acc}}\tilde{E}_{\rm x}-\alpha_{\rm C}\frac{n_{\rm e}}{n_{\rm H}}\tilde{\varepsilon}^{-0.94} \right) \tilde{h}\right \} \nonumber \\
&=& \tilde{Q}_{\rm i,CR}+\tilde{Q}_{\rm i,2nd}+\tilde{Q}_{\rm ex,2nd},
\label{eq:boltzmann3}
\end{eqnarray}
where the resistive electric field, $\tilde{E}_{\rm x}$, the numerical factor, $\alpha_{\rm C}$, and the three terms on the right hand side are  given by 
\begin{eqnarray}
\tilde{E}_{\rm x} &=& \frac{c}{V_{\rm CR}}\int \tilde{h}(\tilde{\varepsilon}_0)d\tilde{\varepsilon}_0 - 1, \\
\alpha_{\rm C} &=& 60.6~{\rm or}~47.6 ~ \text{( for H or ${\rm H}_2$, respectively) }, \\
\tilde{Q}_{\rm i,CR}  &=& \left\{ \frac{v}{c}\frac{d\tilde{\sigma}_{\rm i,p}}{d\tilde{\varepsilon}}(\tilde{\varepsilon},\tilde{\varepsilon}_0)  \right\}_{\tilde{\varepsilon}_0=m_{\rm p}c^2/I_{\rm ion}} ,\\
\tilde{Q}_{\rm i,2nd} &=& \int \frac{d\tilde{\sigma}_{\rm i,e}}{d\tilde{\varepsilon}}(\tilde{\varepsilon},\tilde{\varepsilon}_0) \tilde{h}(\tilde{\varepsilon}_0)d\tilde{\varepsilon}_0 -\tilde{\sigma}_{\rm i,e}(\tilde{\varepsilon})\tilde{h}(\tilde{\varepsilon}),\\
\tilde{Q}_{\rm ex,2nd} &=&  \sum_{k}  \tilde{\sigma}_{{\rm ex},k}(\tilde{\varepsilon}+\tilde{\Delta}_k)\tilde{h}(\tilde{\varepsilon}+\tilde{\Delta}_k) - \tilde{\sigma}_{{\rm ex},k}(\tilde{\varepsilon})\tilde{h}(\tilde{\varepsilon}),~~~~
\end{eqnarray}
All of the above cross sections are normalized by $4\pi a_0^2$ and $\tilde{\Delta}_k=\Delta_k/I_{\rm ion}$.
Then, there are three key parameters in the normalized equation system, $l_{\rm ion}/l_{\rm acc}, n_{\rm e}/n_{\rm H}$, and $V_{\rm CR}/c$. 
$l_{\rm ion}/l_{\rm acc}$ represents the strength of secondary electron acceleration by the initial resistive electric field. 
A larger value of $l_{\rm ion}/l_{\rm acc}$ implies stronger acceleration. 
The electron fraction, $n_{\rm e}/n_{\rm H}$, controls the strength of Coulomb loss, which is the most important parameter 
that determines the secondary electron spectrum deep in the cold region. 
The CR streaming velocity, $V_{\rm CR}/c$, controls the number of CRs. 
A smaller $V_{\rm CR}/c$ means a larger number density of CRs, which leads to a faster generation of secondary electrons.

The boundary condition at $\tilde{x}=0$ is given by $\tilde{h}=0$ and $\tilde{E}_{\rm x}=-1$ because no secondary electrons are generated at $\tilde{x}=0$. 
We assume that the cold region consists of hydrogen molecules in this work. 
Then, $I_{\rm ion}= 15.4~{\rm eV}, \alpha_{\rm C}=47.6$ and $n_{\rm H}=2n_{\rm H_2}$, where $n_{\rm H_2}$ is the number density of hydrogen molecules. 
All the cross sections of hydrogen molecule are taken from \citet{padovani22}.

\section{Results}
\label{sec4}
\subsection{Results for the fiducial parameters}
\begin{figure*}
\centering
\gridline{\fig{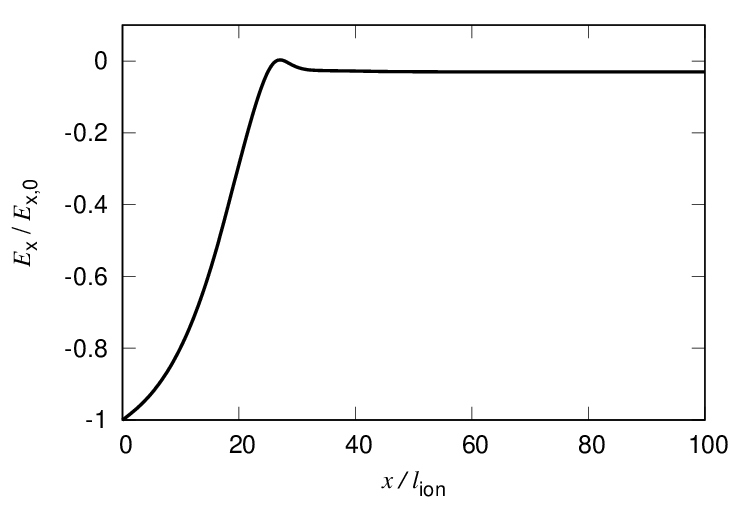}{0.3\textwidth}{}
              \fig{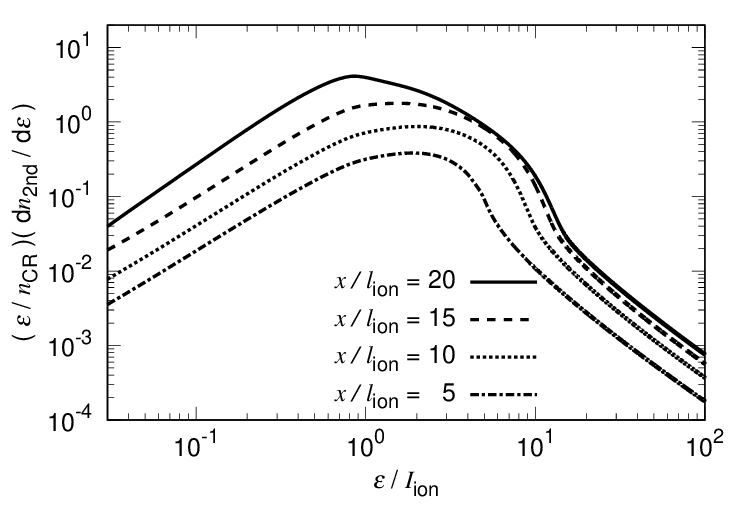}{0.3\textwidth}{}
              \fig{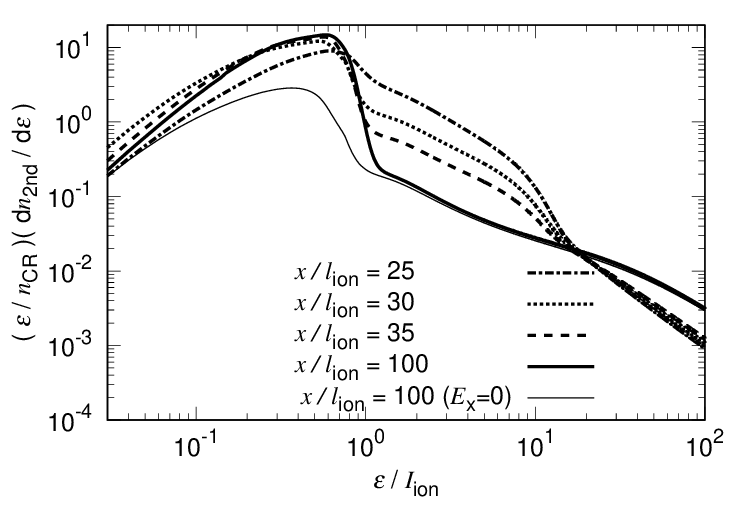}{0.3\textwidth}{}}
\caption{Results for the fiducial parameters, $l_{\rm ion}/l_{\rm acc}=1,n_{\rm e}/n_{\rm H}=10^{-4}, V_{\rm CR}/c=10^{-2}$. The left panel shows the spatial development of the resistive electric field. The middle panel shows spectra at $x/l_{\rm ion}=5, 10, 15, 20$. The right panel shows spectra at $x/l_{\rm ion}=25, 30, 35, 100$. The thin solid line in the right panel shows the spectrum at $x/l_{\rm ion}=100$ for no acceleration by the resistive electric field.}
\label{fig1}
\end{figure*}
\begin{figure}
\centering
\plotone{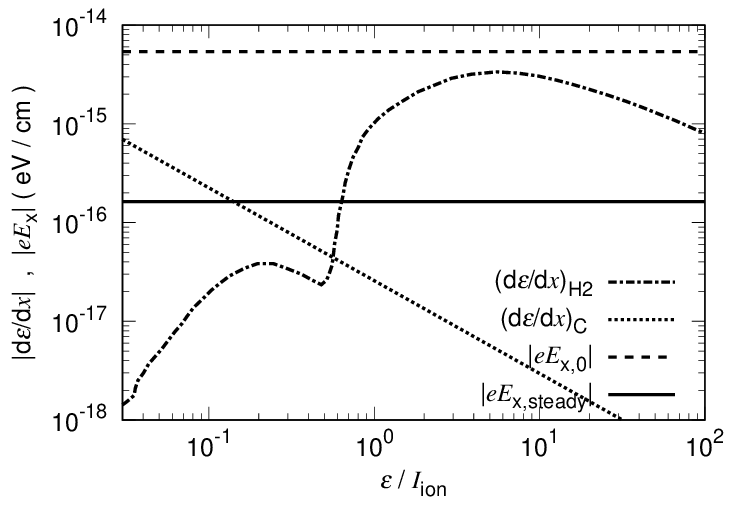}
\caption{Energy loss rates and acceleration rate by the resistive electric field for the fiducial parameters ($l_{\rm ion}/l_{\rm acc}=1,n_{\rm e}/n_{\rm H}=10^{-4}, V_{\rm CR}/c=10^{-2}$) and $n_{\rm H}=1~{\rm cm}^{-3}$.}
\label{fig2}
\end{figure}
First of all, we show simulation results for the fiducial parameters, $l_{\rm ion}/l_{\rm acc}=1,n_{\rm e}/n_{\rm H}=10^{-4}, V_{\rm CR}/c=10^{-2}$. 
The strength of the resistive electric field decreases with $x$ and eventually reaches a finite value (left panel in Figure~\ref{fig1}). 
From the left boundary at $x=0$, secondary electrons are initially injected through CR ionization and accelerated by the resistive electric field. 
When the number of secondary electrons increases to some level, new secondary electrons are mainly injected through ionization by secondary electrons, that is, the electron avalanche happens. 
The current of secondary electrons increases, but the return current of thermal electrons decreases, resulting in a decrease in the resistive electric field. 
Finally, the resistive electric field has a finite strength in a quasi-steady state.

The middle and right panels in Figure~\ref{fig1} show the spatial development of the secondary electron spectrum. 
In the electron avalanche phase (middle panel), the number of secondary electrons exponentially increases and the secondary electrons are accelerated to about $\varepsilon/I_{\rm ion} \sim 10$ by the resistive electric field.
Once the strength of the resistive electric field becomes sufficiently small (right panel), the accelerated secondary electrons lose their energy by ionization and excitation. 
In a quasi-steady state, the acceleration by the weakened resistive electric field is balanced by the energy loss due to ionization and excitation. 
As one can see, compared to the case where the acceleration due to the resistive electric field is ignored (thin sold line in the right panel), 
the resistive electric field significantly enhances the number of secondary electrons with $\varepsilon/I_{\rm ion}\sim 0.6$. 
These secondary electrons can excite hydrogen molecule and ionize atomic carbon, so that the CR discharge would enhance molecular line emissions and ionization of atomic carbon. 
In the fiducial parameter set, the number of secondary electrons with $\varepsilon/I_{\rm ion}> 1$ is significantly enhanced by the CR discharge around $x/l_{\rm ion}\sim 20$, but not strongly enhanced in the quasi-steady state ($x/l_{\rm ion}>35$).

The solid line in the right panel of Figure~\ref{fig3} shows the ionization rate by the secondary electrons for the fiducial parameters. 
As expected from the spectral evolution, the ionization rate is strongly enhanced around $x/l_{\rm ion}\sim 20$ and not enhanced in the quasi-steady state ($x/l_{\rm ion}>35$). 
This is because secondary electrons that can ionize the hydrogen gas are not accelerated by the resistive electric field at the quasi-steady state. 
In the recent standard model in which the resistive electric field is zero, the ratio of the ionization rate by secondary electrons to that by primary CRs is $\zeta_{\rm 2nd}/\zeta_{\rm CR}\approx 1$ \citep{ivlev21}, but $\zeta_{\rm 2nd}/\zeta_{\rm CR}\approx 0.5$ in this work. 
The ionization rate by the secondary electrons gradually increases in the region of $x/l_{\rm ion}>100$ in this work. 
This is because the cooling length of the secondary electrons with $\varepsilon/I_{\rm ion}\sim 10^2$ is larger than our simulation region ($0\le x/l_{\rm ion} \le 500$), that is, the number of secondary electrons around $\varepsilon/I_{\rm ion}\sim 10^2$ still increases with the distance from the left boundary ($x=0$) due to ionization by primary CRs. 
Therefore, the difference in $\zeta_{\rm 2nd}/\zeta_{\rm CR}$ is due to the limitation of spatial and energy ranges in this work.

To understand what happens in the quasi-steady state ($x/l_{\rm ion}>35$), 
we plot energy loss rates and acceleration rate by the resistive electric field in Figure~\ref{fig2}. 
The dot-dashed and dotted curves show the energy loss rates due to interaction with hydrogen molecules and thermal electrons, respectively. 
The dashed and solid horizontal lines show the acceleration rate by the resistive electric field at $x/l_{\rm ion}=0$ and $100$, respectively. 
All secondary electrons are initially accelerated by the resistive electric field because the acceleration rate is larger than the two loss rates. 
As discharge goes on, the current by thermal electrons becomes small, so that the resistive electric field becomes small. 
In the quasi-steady state, very low energy secondary electrons ($\varepsilon/I_{\rm ion}\lesssim 0.1$) are not accelerated and lose their energy by Coulomb interaction. 
In the energy region of $0.1\lesssim\varepsilon/I_{\rm ion}\lesssim 0.6$, secondary electrons can be accelerated, but above this energy range ($\varepsilon/I_{\rm ion} \gtrsim 0.6$), again secondary electrons cannot be accelerated. 
They lose their energy by excitation and ionization processes. 
As you can see, big difference of the secondary electron spectrum appears in the energy range where the electron can be accelerated.  
Secondary electrons in this energy range move to the high energy range, but higher energy electrons move to the lower energy range. 
Hence, the spectral peak is located at $\varepsilon/I_{\rm ion}\approx 0.6$. 
For higher energies, acceleration does not play an important role, so that the resistive electric field does not affect on the secondary electron spectrum. 
If the resistive electric field is larger than the value at the quasi steady state, more secondary electrons are accelerated, so that the electric current by the secondary electrons becomes large and the current by thermal electrons becomes small. 
Then, the electric field becomes small. 
Therefore, the quasi-steady state spectrum of secondary electrons is formed by a balance of the weak acceleration and the energy loss due to excitation and ionization of hydrogen atoms or molecules. 
The initial strength of the resistive electric field is expected not to be important for the steady state. 
That is, the steady state is expected to be controlled by energy loss processes in the cold gas.

\subsection{Dependence of $l_{\rm ion}/l_{\rm acc}$}
\begin{figure*}
\centering
\gridline{\fig{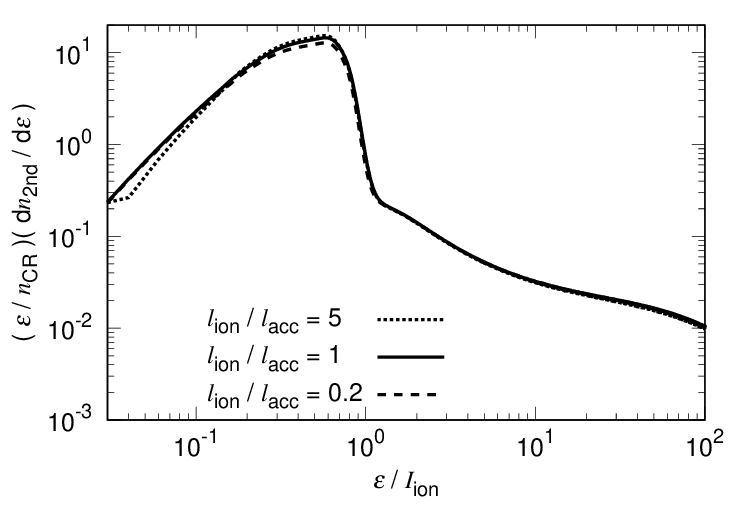}{0.3\textwidth}{}
              \fig{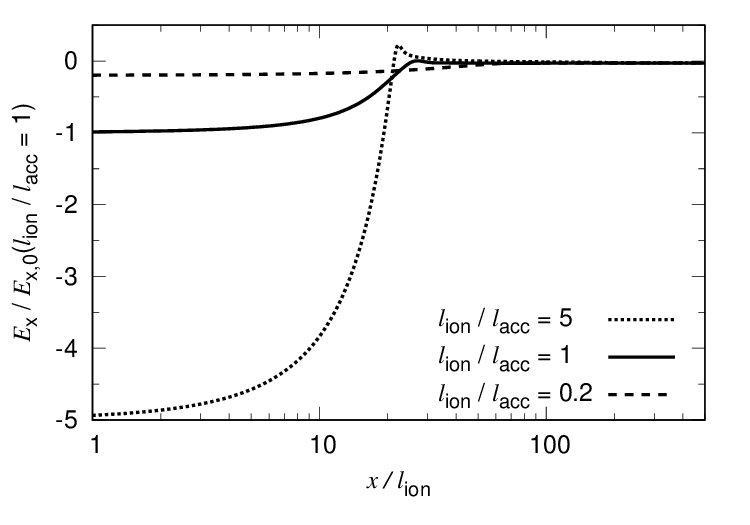}{0.3\textwidth}{}
              \fig{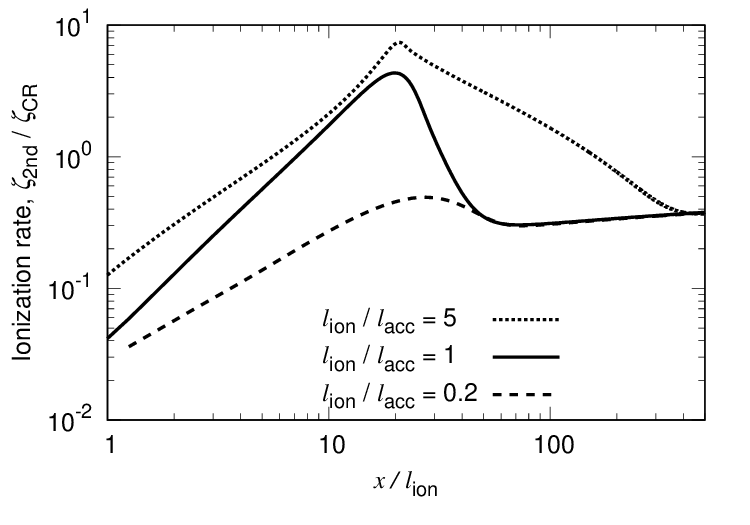}{0.3\textwidth}{}}
\caption{Simulation results for $ l_{\rm ion}/l_{\rm acc}=0.2, 1, 5$ and $n_{\rm e}/n_{\rm H}=10^{-4}, V_{\rm CR}/c=10^{-2}$. The left, middle, and right panels show spectra at $x/l_{\rm ion}= 500$, the spatial evolution of the resistive electric field and the ionization rate by secondary electrons.}
\label{fig3}
\end{figure*}
To study the dependence of the initial resistive electric field, 
we perform some simulations for different values of $l_{\rm ion}/l_{\rm acc}$, keeping other parameters fixed. 
Figure~\ref{fig3} shows the energy spectra of secondary electrons at $x/l_{\rm ion}=500$ (left), the spatial development of the resistive electric field (middle) and ionization rate by secondary electrons (right) for $l_{\rm ion}/l_{\rm acc}=0.2,1,5$ and $n_{\rm e}/n_{\rm H}=10^{-4}$, and $V_{\rm CR}/c=10^{-2}$. 
As one can see, the steady state does not depend on $l_{\rm ion}/l_{\rm acc}$ as long as $l_{\rm ion}/l_{\rm acc}$ is not too small. 
For the larger $l_{\rm ion}/l_{\rm acc}$ case, secondary electrons are quickly accelerated to higher energies before the resistive electric field becomes small. 
Since the cooling length for higher energy electrons is longer, it takes a longer distance for them to reach a steady state.

\subsection{Dependence of $n_{\rm e}/n_{\rm H}$}
\begin{figure*}
\centering
\gridline{\fig{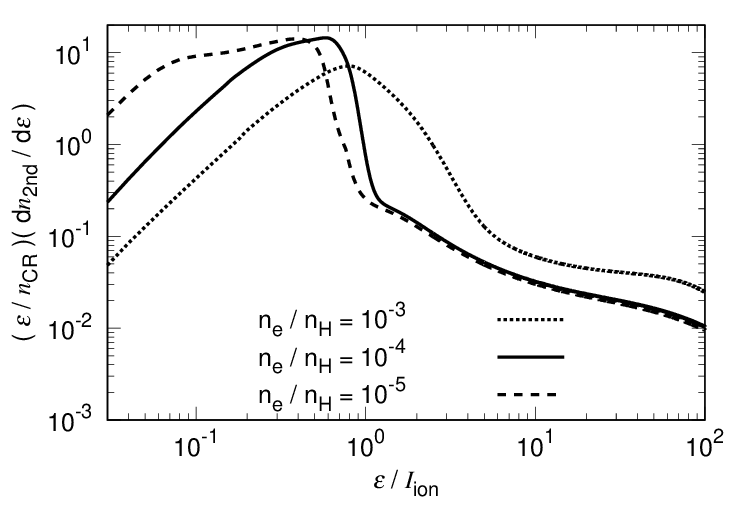}{0.3\textwidth}{}
              \fig{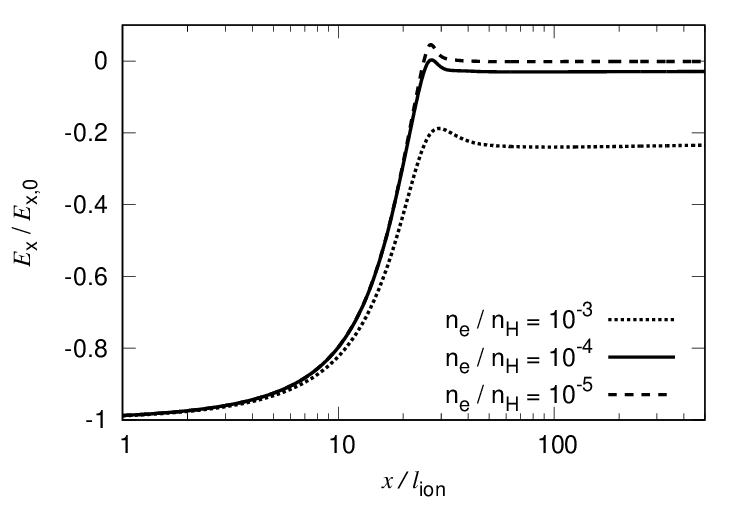}{0.3\textwidth}{}
              \fig{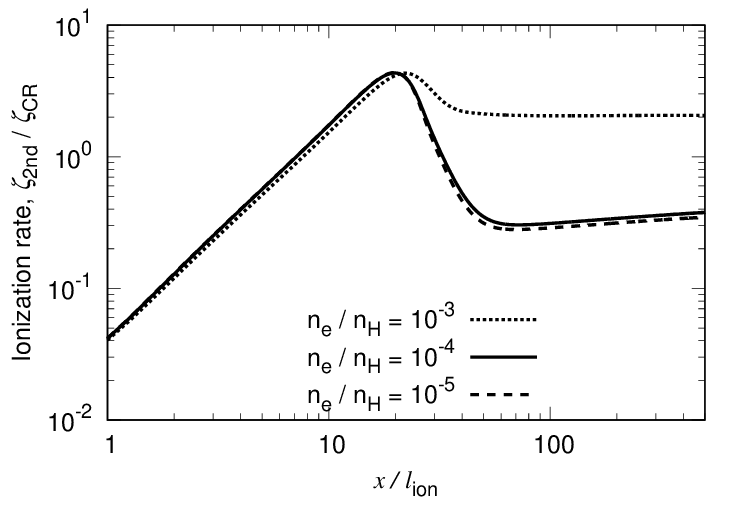}{0.3\textwidth}{}}
\caption{Same as Figure~\ref{fig3}, but for $ n_{\rm e}/n_{\rm H}=10^{-5}, 10^{-4}, 10^{-3}$ and $l_{\rm ion}/l_{\rm acc}=1, V_{\rm CR}/c=10^{-2}$.}
\label{fig4}
\end{figure*}
Next, we investigate dependence of the electron fraction in the cold gas, keeping other parameters fixed.
Figure~\ref{fig4} shows the same as Figure~\ref{fig3} but for $n_{\rm e}/n_{\rm H}=10^{-5}, 10^{-4}, 10^{-3}$ and $l_{\rm acc}/l_{\rm ion}= 1$, and $V_{\rm CR}/c=10^{-2}$. 
For a higher electron fraction, the number of low-energy secondary electrons is smaller, but the number of high-energy secondary electrons is larger, resulting in the larger resistive electric field and higher ionization rate. 
Since the Coulomb loss increases as the electron fraction increases, the low-energy secondary electrons lose their energy more quickly. 
Then, the number and electric current of low-energy secondary electrons become smaller but the resistive electric field is larger. 
As a result, secondary electrons that can ionize the cold gas are accelerated more. 
For the electron fraction of $10^{-3}$, the ionization rate by the secondary electrons is about five times larger than one of the standard model. 
Therefore, the electron fraction is a crucial parameter to control the quasi-stead state after the CR discharge. 
It should be noted that all the parameters of the background cold gas are fixed during the CR discharge in this study. 
The CR discharge might change the electron fraction and gas temperature, which might affect the evolution of the CR discharge itself. 
These back reaction of the CR discharge on the cold gas should be addressed in the future.

\subsection{Dependence of $V_{\rm CR}/c$}
\begin{figure*}
\centering
\gridline{\fig{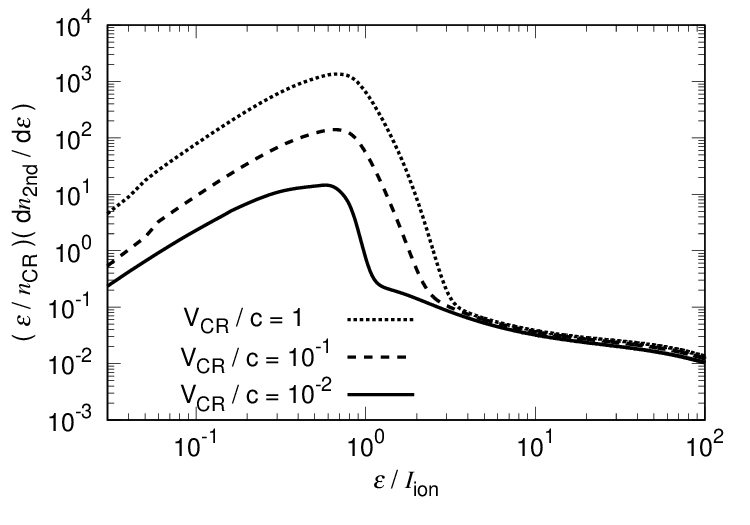}{0.3\textwidth}{}
              \fig{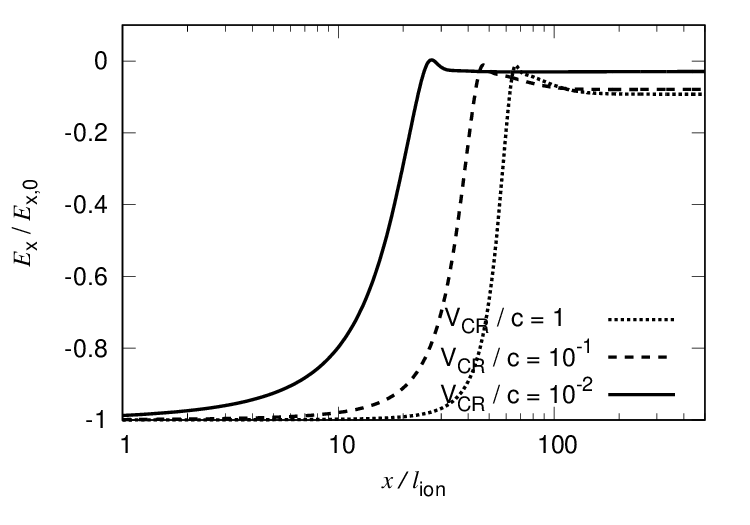}{0.3\textwidth}{}
              \fig{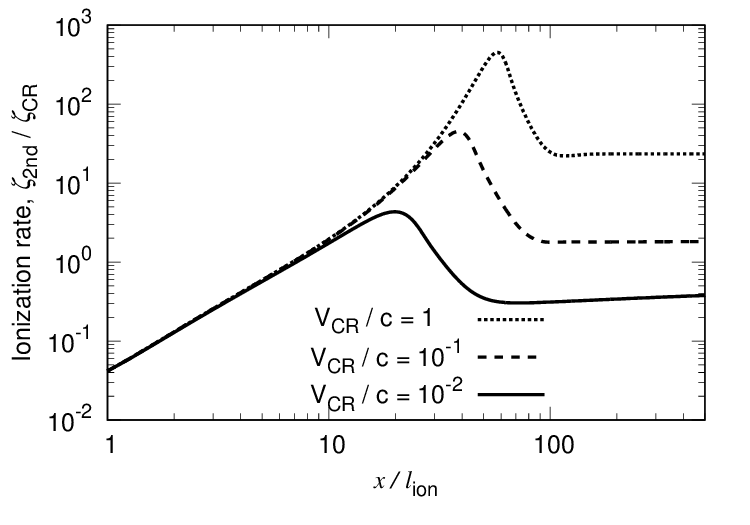}{0.3\textwidth}{}}
\caption{Same as Figure~\ref{fig3}, but for $V_{\rm CR}/c=10^{-2}, 10^{-1}, 1$ and $l_{\rm ion}/l_{\rm acc}=1, n_{\rm e}/n_{\rm H}=10^{-4}$.}
\label{fig5}
\end{figure*}
Finally, we investigate dependence of the drift velocity of primary CRs in the cold gas, keeping other parameters fixed.
Figure~\ref{fig5} shows the same as Figure~\ref{fig3} but for $V_{\rm CR}/c=10^{-2}, 10^{-1}, 1$ and $l_{\rm acc}/l_{\rm ion}= 1, n_{\rm e}/n_{\rm H}=10^{-4}$. 
Changing the drift velocity while keeping the CR flux means changing the number density of CRs. 
The number density of CR is larger for the lower drift velocity, so that secondary electrons are injected by the primary CRs more rapidly. 
As a result, the electron avalanche occurs more rapidly and the system reaches the quasi-steady state more rapidly. 
Since energy loss processes of secondary electrons in the cold gas do not depend on the drift velocity of CRs, 
in the quasi-steady stage, the resistive electric field and ionization rate by secondary electrons are expected to weakly depend on the drift velocity of CRs. 
This expectation is consistent with our simulation results, but it should be noted that the vertical axises of the left and right panels are normalized by the number density of primary CRs. 
As you can see in the left and right panels of Figure~\ref{fig5}, the absolute values of ionization rate by the secondary electrons and the number of low-energy secondary electron almost do not depend on the drift velocity of CRs.  
On the other hand, normalized spectra of secondary electrons in the high-energy region ($\varepsilon/I_{\rm ion}\gtrsim 1$) are almost the same. 
Since the high-energy secondary electrons are mainly produced by ionization by primary CRs, the number of the high-energy secondary electrons normalized by the number of CRs does not depend on the drift velocity of CRs. 
Therefore, the drift velocity of CRs is not a crucial parameter in terms of the absolute value of the ionization rate by secondary electrons in the quasi-steady state.

\section{Discussion}
\label{sec5}
In our previous paper \citep{ohira22}, the evolution of the resistive electric field was studied by neglecting ionization by newly generated secondary electrons, that is, secondary electrons were generated only by primary CR ionization. 
In this work, we have taken into account ionization by the secondary electrons and shown that the secondary electron ionization dominates primary CR ionization around $x/l_{\rm ion}=20$.
For some case (e.g. for $n_{\rm e}/n_{\rm H}=10^{-1}$ or for $V_{\rm CR}/c \gtrsim 10^{-1}$), the secondary electron ionization is dominant even in $x/l_{\rm ion}\gtrsim30$. 
Secondary electrons are generated more quickly by the electron avalanche, so that the strength of the resistive electric field decreases more quickly. 
As a result, the maximum energy of accelerated secondary electrons is smaller than the estimation in our previous paper \citep{ohira22}. 
Even after the quick decay of the resistive electric field ($x/l_{\rm ion}\gtrsim30$), the resistive electric field does not disappear, but has a lower value than one before the CR discharge. 
The weakened resistive electric field can gradually accelerate high-energy secondary electrons  ($\varepsilon/I_{\rm ion}\gg 10^2$) generated by the primary CRs. 
As the accelerated high-energy secondary electrons contribute to the return current, the resistive electric field would decrease further. 
The long range evolution of the CR discharge will be addressed in future work, which is important for understanding the CR ionization deep in molecular clouds.

In this work, we did not take into account effects of magnetic fields and the inelastic scattering between secondary electrons and hydrogen molecules. 
Further more, we assumed that all secondary electrons move to the direction of the CR streaming. 
Although the inelastic scattering does not contribute to the energy loss in the energy range that we are interested in, 
it changes the direction of momentum of secondary electrons. 
Thus, the inelastic scattering makes the momentum distribution of secondary electrons isotropic. 
On the other hand, the resistive electric field accelerates secondary electrons to the direction of the CR streaming. 
Furthermore, a local magnetic field changes motion of the secondary electrons locally. 
Therefore, strictly speaking, we have to solve the time evolution of the momentum distribution in the full six-dimensional phase space, which requires enormous computational effort. 
In addition, Ohm's law may need to be modified to account for the resistivity of secondary electrons due to the elastic scattering. 
Furthermore, interactions with electromagnetic waves could cause some anomalous resistivity, which could enhance the resistive electric field.

Our simulations showed that the CR discharge significantly increases the number of secondary electrons with energy around $\varepsilon \sim 0.6~I_{\rm ion}$. 
These secondary electrons can ionize some atoms (e.g. Carbon) and excite many atoms and molecules. 
Therefore, detailed calculations for these emission lines and comparisons with observational data would provide us a new information about CRs.

CRs would be present even in the early universe at the redshift of $z\approx 20$ \citep{ohira19} and play an important roll in early galaxies and the intergalactic medium (IGM) \citep{miniati11,ohira20,ohira21,yokoyama22,yokoyama23}. 
Since the IGM at the redshift $z\approx 20$ is cold and very dilute ($T\sim 1~{\rm K}$ and $n_{\rm H}\sim 10^{-3}~{\rm cm}^{-3}$) compared with cold gases in the current universe, the ratio of $l_{\rm ion}/l_{\rm acc}$ is expected to be very large and the CR discharge would significantly affect ionization and heating of IGM in the early universe. 
Since future 21 cm observations will be able to access the IGM heating by CRs in the early universe \citep{gessey23}, the CR discharge in the early universe should be addressed in the future. 
\section{Summary}
\label{sec6}
Cold dense regions in the current universe like molecular clouds are ionized mainly by CRs. 
In the standard picture, secondary electrons generated by the CR ionization lose their energy in the dense region, and on average, one secondary electron makes about one secondary electrons before losing their energy \citep{ivlev21}. 
In this work, we have considered streaming CRs which drive the electron return current.
The resistivity of the return current induces a resistive electric field. 
Then, secondary electrons are accelerated by the resistive electric field as long as the CR current is sufficiently large. 
By solving the one-dimensional steady-state Boltzmann equation and Ohm’s law, we have investigated the spatial evolution of the secondary electron spectrum. 
We have showed that the electron avalanche happens, that is, one secondary electrons makes a lot of secondary electrons. 
As a result, the CR discharge increases the ionization and excitation rates, which could explain the high ionization rate that are observed in some molecular clouds. 
A quasi-steady state is eventually realized after the electron avalanche happens. 
We found that the quasi-steady state almost does not depend on the initial resistive electric field, but depends on the electron fraction in the cold dense region. 
The higher electron fraction increases the resistive electric field in the quasi-steady state, so that the number of secondary electrons that can ionize the gas is greater, resulting in a higher ionization rate.

\begin{acknowledgments}
We are grateful to the referee for valuable comments that improved the paper. 
We also thank Diniz Gabriel, Teruaki Enoto, Kazuhiro Nakazawa, and Yuuki Wada for informative discussions on discharge in thunderclouds. 
Numerical computations were carried out on Cray XC50 at Center for Computational Astrophysics, National Astronomical Observatory of Japan. 
This work is supported by JSPS KAKENHI Grant Number JP21H04487 and JP24H01805. 
\end{acknowledgments}


\begin{thebibliography}{}
%
\bibitem[Caselli et al.(1998)]{caselli98}
Caselli, P., Walmsley, C. M., Terzieva, R., \& Herbst, E. 1998, \apj, 499, 234
%
\bibitem[Dalgarno(2006)]{dalgarno06}
Dalgarno, A. 2006, Publ. Nat. Acad. Sci., 103, 12269
%
\bibitem[Gabici(2022)]{gabici22}
Gabici, S. 2022, \aapr, 30, 4
%
\bibitem[Gaches et al.(2021)]{gaches21}
Gaches, B. A. L., Walch, S. \& Lazarian, A. 2021, \apjl, 917, L39
%
\bibitem[Gessey-Jones et al.(2023)]{gessey23} 
Gessey-Jones, T., Fialkov, A., de Lera Acedo, E., Handley, W. J., \& Barkana, R. 2023, \mnras, 526, 4262
%
\bibitem[Fujita et al.(2009)]{fujita09}
Fujita, Y., Ohira, Y.,  Tanaka, S. J. \& Takahara, F. 2009, \apjl, 707, L179
%
\bibitem[Fujita et al.(2010)]{fujita10}
Fujita, Y., Ohira, Y. \& Takahara, F. 2010, \apjl, 712, L153
%
\bibitem[Hayakawa et al.(1961)]{hayakawa61}
Hayakawa, S., Nishimura, S., \& Takayanagi, T. 1961,  \pasj, 13, 184
%
\bibitem[Indriolo \& McCall(2012)]{indriolo12}
Indriolo, N., \& McCall, B. J. 2012, \apj, 745, 91
%
\bibitem[Indriolo et al.(2015)]{indriolo15}
Indriolo, N., Neufeld, D. A., Gerin, M., et al. 2015, \apj, 800, 40
%
\bibitem[Ivlev et al.(2021)]{ivlev21}
Ivlev, A. V., Silsbee, K., Padovani, M., \& Galli, D. 2021, \apj, 909, 107
%
\bibitem[Malkov et al.(2013)]{malkov13}
Malkov, M. A., Diamond, P. H., Sagdeev, R. Z., Aharonian, F. A., \& Moskalenko, I. V. 2013 \apj, 768, 73
%
\bibitem[Malkov et al.(2016)]{malkov16}
Malkov, M. A., Diamond, P. H., \& Sagdeev, R. Z. 2016 \prd, 94, 063006
%
\bibitem[Miniati \& Bell(2011)]{miniati11}
Miniati, F., \& Bell, A. R. 2011, \apj 729, 73.
%
\bibitem[Ohira et al.(2011)]{ohira11}
Ohira, Y., Murase, K., \& Yamazaki, R. 2011, \mnras, 410, 1577
%
\bibitem[Ohira \& Murase(2019)]{ohira19}
Ohira, Y., \& Murase, K. 2019, \prd, 100, 061301(R)
%
\bibitem[Ohira(2020)]{ohira20}
Ohira, Y. 2020, \apjl, 896, L12
%
\bibitem[Ohira(2021)]{ohira21}
Ohira, Y. 2021, \apj, 911, 26 
%
\bibitem[Ohira (2022)]{ohira22}
Ohira, Y. 2022, \apj, 929, 106
%
\bibitem[Padovani et al.(2016)]{padovani16}
Padovani, M., Marcowith, A., Hennebelle, P., Ferri{\'e}re, K. 2016, \aap, 590, A8 
%
\bibitem[Padovani et al.(2020)]{padovani20}
Padovani, M., Ivlev, A. V., Galli, D., Offner, S. S. R., Indriolo, N., Rodgers-Lee, D., Marcowith, A., Girichidis. P., Bykov, A. M., Kruijssen, J. M. D. 2020, \ssr, 216, 29
%
\bibitem[Padovani et al.(2022)]{padovani22}
Padovani, M. et. al. 2022, \aap, 658, A189
%
\bibitem[Phan et al.(2023)]{phan23}
Phan, V. H. M., Recchia, S., Mertsch, P., \& Gabici, S. \prd, 2023
%
\bibitem[Silsbee \& Ivlev (2020)]{silsbee20}
Silsbee, K., \& Ivlev, A. V. 2020, \apjl, 902, L25
%
\bibitem[Spitzer \& Tomasko(1968)]{spitzer68}
Spitzer, L., \& Tomasko, M. G. 1968, \apj, 152, 971
%
\bibitem[Swartz et al.(1971)]{swartz71}
Swartz, W. E., Nisbet, J. S., \& Green, A. E. S. 1971, \jgr, 76, 8425
%
\bibitem[Torres et al.(2008)]{torres08}
Torres, D. F., Rodriguez Marrero, A. Y. \& de Cea Del Pozo, E. 2008, \mnras, 387, L59
%
\bibitem[Yang \& Wang(2020)]{yang20}
Yang, R.-Z., \& Wang, Y. 2020, \aap, 640, A60
%
\bibitem[Yokoyama \& Ohira (2022)]{yokoyama22}
Yokoyama, S. L. \& Ohira, Y. 2022, \mnras, 515, 5467
%
\bibitem[Yokoyama \& Ohira (2023)]{yokoyama23}
Yokoyama, S. L. \& Ohira, Y. 2023, \mnras, 523, 3671
%
\end{thebibliography}
\end{document}